\documentclass{elsart}
\usepackage{graphicx,amssymb}
\journal{Eur. Phys. J. C}

\begin{document}
\thispagestyle{empty}

\def\etal{{\sl et al.}}
\def\retal{{\sl I DR.}}
\newcommand{\be}{\begin{eqnarray}}
\newcommand{\ee}{\end{eqnarray}}
\newcommand{\bls}{\Biggl[}
\newcommand{\brs}{\Biggr]}
\newcommand{\No}{$\numbersign$ }
\def\th{{th}}

\newcommand{\dg}{$\Delta g$}
\newcommand{\dhh}{$\Delta h$}
\newcommand{\dk}{$\Delta k$}
\newcommand{\pgt}{$P(>Z)$}
\newcommand{\chis}{$\chi^{2} $}
\newcommand{\chin}{$\chi^{2} / NDF$}
\newcommand{\my}{$<Y>$}
\newcommand{\mx}{$<X>$}
\newcommand{\kp}{$K^{+}$}
\newcommand{\km}{$K^{-}$}
\newcommand{\kpm}{$K^{\pm}$}
\newcommand{\kdc}{$K^{\pm}\rightarrow \pi^{\pm} \pi^{0} \pi^{0}$}
\newcommand{\kpdc}{$K^{+} \rightarrow \pi^{+}\pi^{0}\pi^{0} $}
\newcommand{\kmdc}{$K^{-} \rightarrow \pi^{-}\pi^{0}\pi^{0} $}
\newcommand{\knl}{$K^{\pm} \rightarrow 3\pi $}
\newcommand{\kdccc}{$K^{\pm} \rightarrow \pi^{\pm}\pi^{+}\pi^{-}$}
\newcommand{\kdcg}{$K^{\pm} \rightarrow \pi^{\pm}\pi^{0}\ \gamma $}
\newcommand{\ac}{$A_{g}$}
\newcommand{\aoo}{$A_{g}^{0}$}
\newcommand{\agc}{$A_{gC}$}
\newcommand{\agn}{$A_{gN}$}
\newcommand{\adg}{$| \Delta g ( \kdc\ )|$}
\newcommand{\re}{$( \epsilon' / \epsilon) $}
\newcommand{\rre}{$Re( \epsilon' / \epsilon) $}

\begin{frontmatter}

\author{G.A.~Akopdzhanov},
\author{V.B.~Anikeev},
\author{V.A.~Bezzubov},
\author{S.P.~Denisov},
\author{A.A.~Durum},
\author{Yu.V.~Gilitsky},
\author{S.N.~Gurzhiev},
\author{V.M.~Korablev},
\author{V.I.~Koreshev},
\author{A.V~Kozelov\thanksref{cora}},
\author{E.A.~Kozlovsky},
\author{V.I.~Kurbakov},
\author{V.V.~Lipaev},
\author{V.A.~Onuchin},
\author{A.M.~Rybin},
\author{\fbox{Yu.M.~Sapunov}},
\author{A.A.~Schukin},
\author{M.M.~Soldatov},
\author{D.A.~Stoyanova},
\author{K.I.~Trushin},
\author{I.A.~Vasilyev},
\author{V.I.~Yakimchuk},
\author{S.A.~Zvyagintsev}
\address{State Research Center 
Institute for High Energy Physics, Protvino, 
Moscow oblast, 142281 Russia}
%
\thanks[cora]{Corresponding author. {\it E-mail address:} kozelov@mx.ihep.su}

\title{Measurements of the Charge Asymmetry of the Dalitz Plot Parameters 
for $K^{\pm}\to\pi^{\pm}\pi^0\pi^0$ Decays }



\begin{abstract}

The charge asymmetry of the $g$, $h$, and $k$ Dalitz plot 
parameters for \kdc\ decays has been measured with 35~GeV/c 
hadron beams at the 70~GeV IHEP accelerator. 
The $g$, $h$, and $k$ values obtained appear to be identical for $K^{\pm}$ decays within the errors quoted.
In particular, the charge asymmetry $A_g = (g^+ - g^-)/(g^+ + g^-)$
of the slope $g$ is equal to $(0.2 \pm 1.9)\cdot 10^{-3}$.
\end{abstract}

\end{frontmatter}

\section{INTRODUCTION}

Observation of direct CP violation in neutral kaon decays 
\cite{na48.1,na48.2,ktev.1} 
gives arguments to search for a similar effect
in charged kaon decays. For example, this effect can manifest itself 
as a charge asymmetry of the Dalitz plot parameters for  
\kdc\ decays. 
These parameters are coefficients in a series expansion 
of the squared module of the matrix element \cite{pdg}:
\begin{eqnarray} 
|M(u,v)|^{2} \propto 1 + gu + hu^{2} + kv^{2} , 
\end{eqnarray}
where $u$ and $v$ are invariant variables.

Theoretical estimates of the charge asymmetry of the Dalitz plot 
slope $g$ for  \kdc\ decays are uncertain and 
range from $10^{-6}$ to $10^{-3}$ 
\cite{bel.2,ambro,isido,gamiz}. 
In the majority of the experiments, only $g^+$ or $g^-$ was measured \cite{pdg,istra}. From these studies it follows that 
$\Delta g = g^{+} - g^{-} = 0.066 \pm 0.017 $. 
It is very unlikely to expect direct CP violation at this level, 
and one can assume that the above mentioned difference is due to the underestimation of the systematic uncertainties.

$K \rightarrow 3\pi$ decays have been studied simultaneously for both 
$K^+$ and $K^-$ mesons in \cite{ford,smith,prelim}. 
Ford {\it et al.} \cite{ford} found $ A_g = -0.0070 \pm 0.0053$ for
$K^{\pm}\rightarrow\pi^{\pm}\pi^{\pm}\pi^{\mp}$ decays.
In the experiment \cite{smith}, $A_g = 0.0019 \pm 0.0123$ was measured 
for \kdc\ decays. 
Preliminary analysis of our experimental data \cite{prelim} based on a fraction of statistics  
yielded $A_g = -0.0003 $ with 
a statistical error of 0.0025 and 
a systematical uncertainty below 0.0015.       
In this paper we report our final results on the
charge asymmetry of the Dalitz plot parameter measurements.

\section{EXPERIMENTAL SETUP}

\begin{figure}
\begin{center}
\includegraphics*[angle=90,scale=0.5]{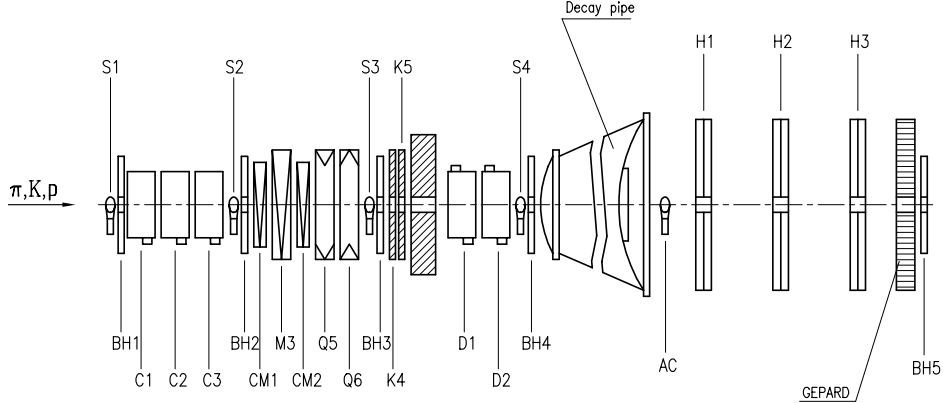}
\end{center}
\caption{
Experimental layout: M -- magnets, Q -- quadrupoles, CM -- tuning magnets,
K -- collimators, S -- scintillation counters, C,D -- threshold and differential
$\mathrm{\check C}$herenkov counters, 
BH -- beam hodoscopes, AC -- anticoincidence counter,
H -- scintillation hodoscopes, GEPARD -- electromagnetic calorimeter.}
\label{fig.setup}
\end{figure}

\begin{figure}
\begin{center}
\includegraphics*[scale=0.5,bb=50 30 567 567 ]{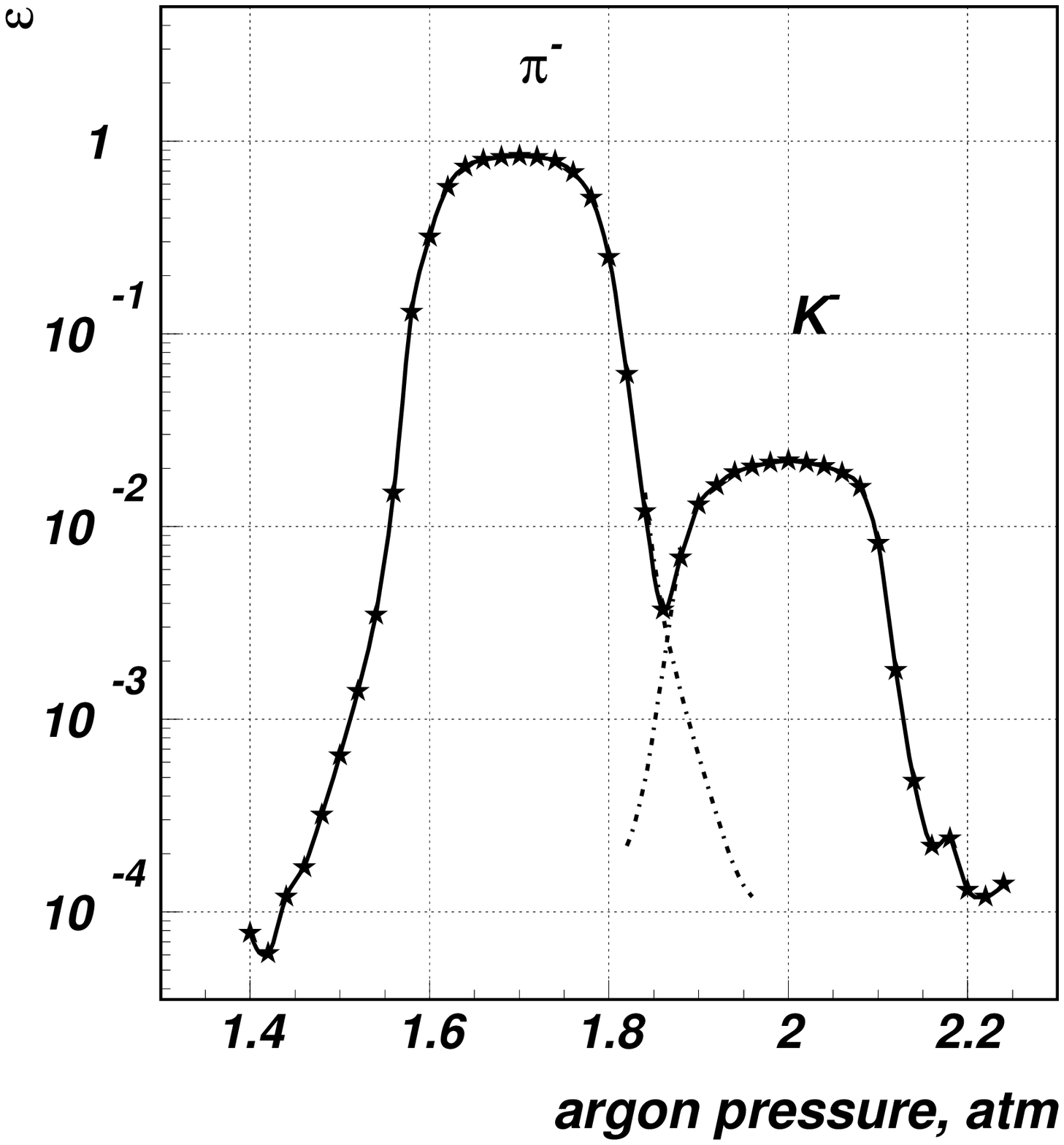}
\end{center}
\caption{
Efficiency of the differential $\mathrm{\check C}$herenkov 
counters vs argon pressure.
}
\label{fig.kpeak}
\end{figure}

The experiment was carried out with 
the TNF-IHEP facility \cite{proposal}
~(Fig.~\ref{fig.setup}) at the 70~GeV IHEP accelerator. 
Unseparated 35~GeV/c positive and negative hadron beams used for kaon decay studies are produced by 70~GeV protons in the external 30~cm Al target. Scintillation counters S1-S4 and beam hodoscopes BH1-BH4 are used to monitor beam intensity and to measure particle trajectories and beam profiles. 
The typical particle flux was 4$\times$10$^{6}$ per 1.7 second spill.

Kaons are selected with three threshold C1-C3 and two differential D1, D2 gas $\mathrm{\check C}$herenkov counters 
(Fig.~\ref{fig.setup}).
The admixture of unwanted particles under the kaon peak
was substantially below 1\%  (Fig.~\ref{fig.kpeak}). 
The threshold counters are also used to select
10~GeV/c electrons to calibrate GEPARD calorimeter.

About 20\% of kaons decay in the 58.5~m long vacuum pipe 
located downstream of the BH4 hodoscope. The flanges of the vacuum pipe have thin Mylar windows in the path of beam particles. The 3.6~m diameter exit flange
is made of 4~mm thick (0.23~$X_0$) stainless steel.
The probability of a high-energy photon to convert 
into an $e^+e^-$ pair in this flange is equal to 0.16. 

Kaons which pass through the decay pipe are detected by 
the anticoincidence counter AC.
The BH5 beam hodoscope placed behind the calorimeter is used for a high precision measurement of the beam position at the setup end. 
The BH5 hodoscope operates in the counting mode and hence detects 
all beam particles.

\begin{figure}
\begin{center}
\includegraphics*[scale=0.6,angle=90,bb=160 18 579 812 ]{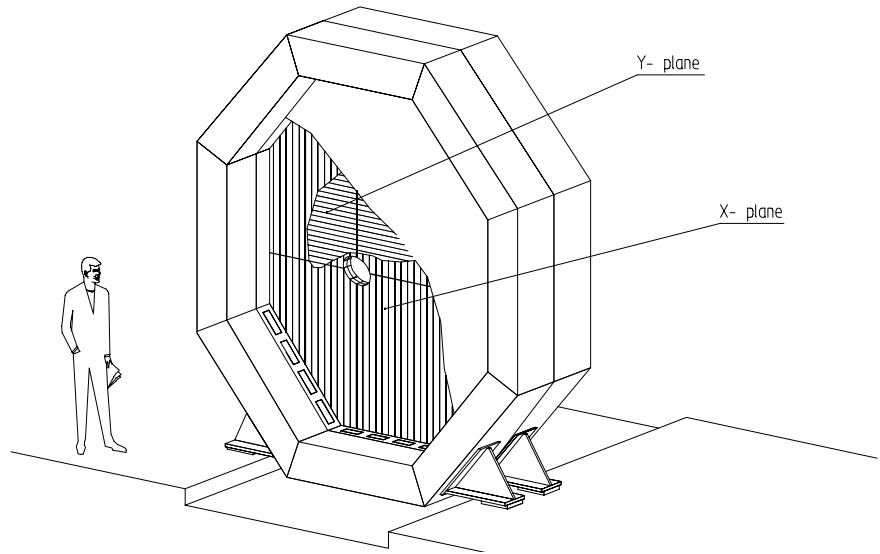}
\end{center}
\caption{
General view of the scintillation hodoscope H.
}
\label{fig.hodo}
\end{figure}

\begin{figure}
\begin{center}
\includegraphics*[scale=0.6,angle=90,bb=100 18 579 700]{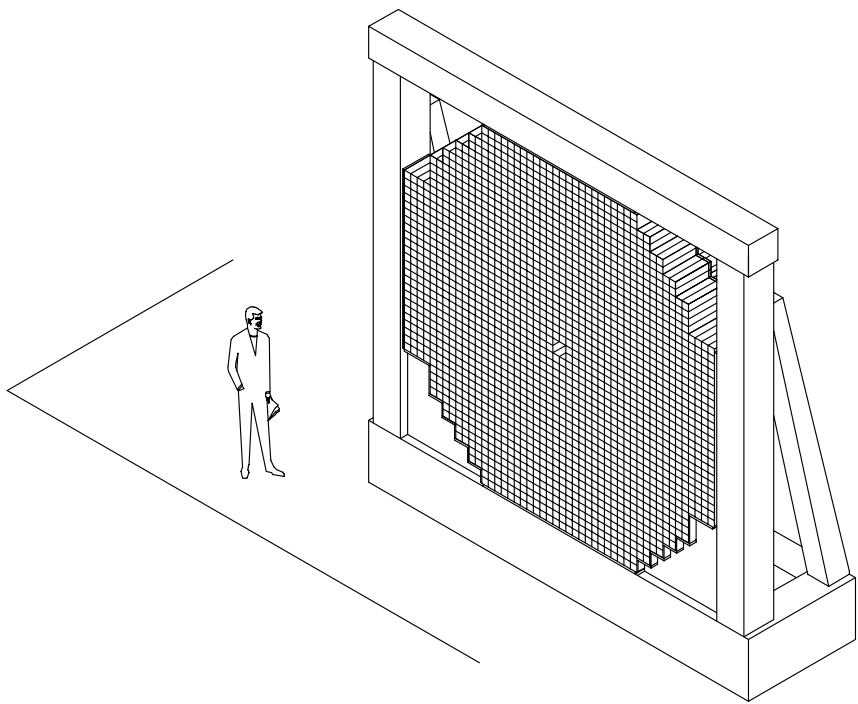}
\end{center}
\caption{
The GEPARD calorimeter.
}
\label{fig.gepard}
\end{figure}

The products of kaon decays are detected by 
three scintillation hodoscopes H1-H3 \cite{hodo} 
and the GEPARD electromagnetic calorimeter. Each hodoscope 
is made of two $X$,$Y$ octagonal planes
with 3.85~m distance between the opposite octagonal sides
(Fig.~\ref{fig.hodo}). 
The plane is divided into half-planes with 256 elements each. The cross section of the hodoscope elements is $14 \times 12$~mm$^2$
and their length varies from 0.7 to 1.8~m. 
Scintillation light is detected by
FEU-84-3 photomultiplier tubes. 

\begin{figure}
\begin{center}
\includegraphics*[scale=0.9,bb=0 0 397 283]{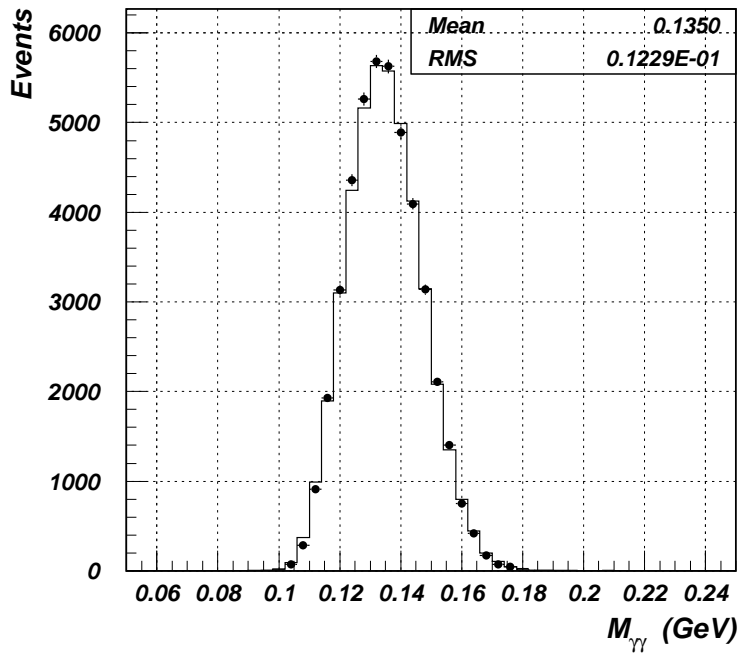}
\end{center}
\caption{
M$_{\gamma\gamma}$ distribution for $K^{\pm}\rightarrow\pi^{\pm}\pi^{0}$ decays.
}
\label{fig.pi0mass}
\end{figure}

The GEPARD is a sampling lead-scintillator calorimeter. It contains 1968 cells with $76 \times 76$~mm$^2$ cross section (Fig.~\ref{fig.gepard}).
Each cell consists of 40 alternating layers of 3~mm Pb and 5~mm scintillator.
The total radiation length is 21~$X_0$.   
Scintillation light is collected onto 
FEU-84-3 photomultiplier tubes using wavelength shifting light guides. 
The GEPARD calorimeter was calibrated by
irradiating each cell with 10~GeV/c electrons at the
beginning of data taking and by using
$K^{\pm}\rightarrow\pi^{\pm}\pi^0$ reconstructed events collected during 
the experiment. Both methods yielded consistent results.  
The $\pi^0$ mass resolution is equal to 12.3~MeV/c$^2$ 
(Fig.~\ref{fig.pi0mass}).

The Level~1 trigger is formed according to the logic formula 
$$T1=S1\cdot S2\cdot S3\cdot S4\cdot (D1+D2)\cdot\overline{C1}
\cdot\overline{C2}\cdot\overline{C3}\cdot\overline{AC}.$$

The Level~2 trigger uses information about energy deposition in the GEPARD calorimeter \cite{trig}.
For this purpose the calorimeter is divided into 16 trigger elements. 
The Level 2 trigger is formed if the energy deposition exceeds 0.8~GeV
in at least three trigger channels.

\begin{figure}
\begin{center}
\includegraphics*[scale=1.,bb=50 0 510 200 ]{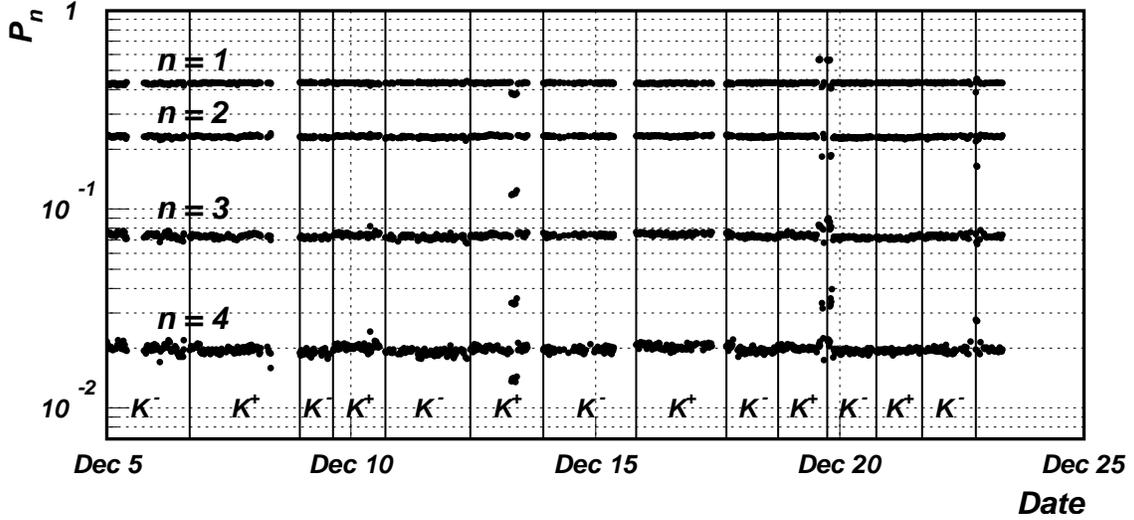}
\end{center}
\caption{
Time variations of the probability $P_{n}$ to reconstruct $n$ tracks ($P_{n}$
is averaged over $\sim$10$^{5}$ events).
}
\label{fig.stab_tracks}
\end{figure}

\begin{figure}
\begin{center}
\includegraphics*[scale=1.,bb=50 0 510 200 ]{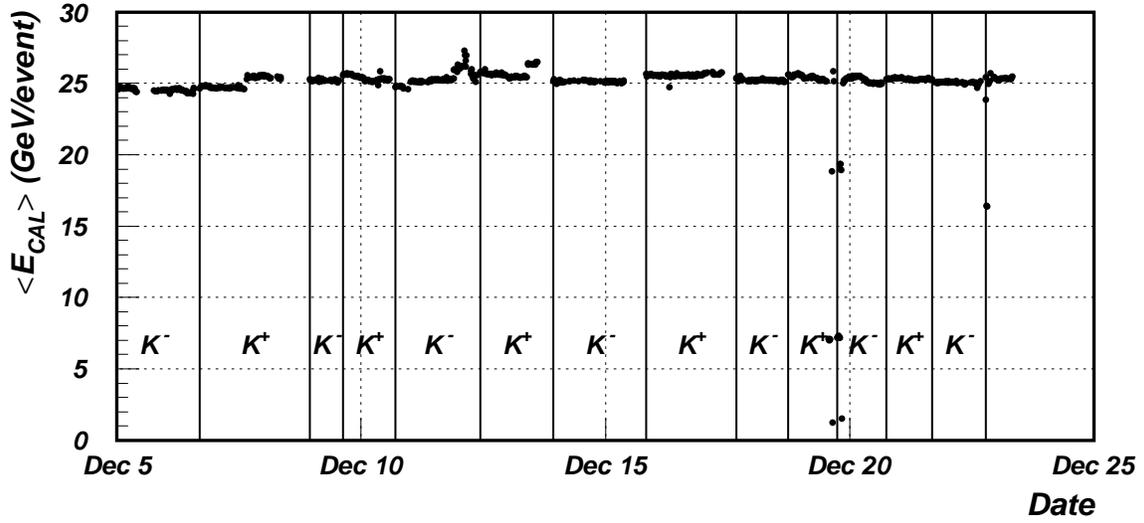}
\end{center}
\caption{
Time variation of the mean energy per event detected in the GEPARD
calorimeter ($E_{Cal}$ is averaged over $\sim$10$^{5}$ events).
}
\label{fig.stab_energy}
\end{figure}

The stability of the beam and detector parameters was carefully
monitored during the data collection. 
To reduce the systematical uncertainty in the measurement of 
the charge asymmetry of the Dalitz plot parameters 
the beam polarity was reversed every day.
Figs.~\ref{fig.stab_tracks}-\ref{fig.stab_energy} illustrate the stability of the experimental setup operation.

\section{EVENT RECONSTRUCTION AND SELECTION CRITERIA}

The \kdc\ event selection starts by finding  
energy clusters in the GEPARD calorimeter.  
The coordinates of the cluster centers and the $X$ and $Y$ coordinates
measured by H1--H3 hodoscopes are used 
in track reconstruction. 
To reduce the combinatorial background, only 
tracks with three or four hits  
in each $X$ and $Y$ projection are selected. 
Then the vertex position of the $K^{\pm}$ decay 
is calculated using the reconstructed tracks.
A track is considered to be associated with a kaon decay if the hypothesis of its intersection with beam axis has a confidence level of 5\% or more and the decay vertex position is inside the fiducial volume of the decay pipe. In addition selected events have
to satisfy one of the following criteria:
\begin{itemize}
\item  five clusters with energies above 1~GeV 
are found and each track is 
associated with one of these clusters; 
\item  four clusters with energies above 1~GeV are found 
and one of the tracks is not associated with 
these clusters. 
\end{itemize}

These criteria are applied because there is a substantial probability 
for the gamma from $\pi^0$  
decays to convert into $e^+e^-$ pair in the exit flange of the decay pipe
(see Section 2), 
and charged pion energy deposition in the 
calorimeter could exceed the threshold value of 1~GeV.

Events passing this preliminary selection 
are subjected to a kinematic fit that allows one   
to resolve ambiguities due to the
combinatorial background 
(for example, to associate one of the tracks with the charged pion) 
and to calculate the Dalitz plot variables.
Altogether 21 measured parameters are used in the fitting procedure:
the energies and the coordinates of four clusters associated with gammas,
the kaon mean energy and the parameters of the kaon and pion tracks.
The parameters of the clusters are corrected for the transverse profile of 
the electromagnetic shower and for the spatial nonuniformity of the calorimeter. The energy of the charged pion is the only unknown parameter.

\begin{figure}
\begin{center}
\includegraphics*[scale=0.9,bb=0 10 340 250]{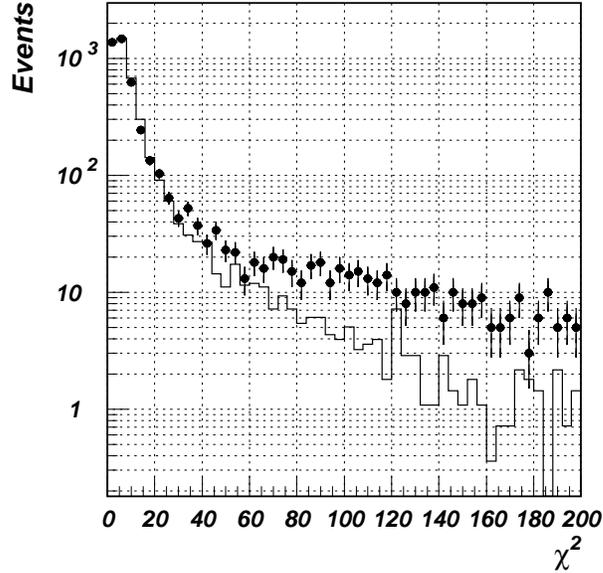}
\end{center}
\caption{
$\chi^{2}$ distribution for \kdc\ decays (histogram -- simulation, circles -- experiment).}
\label{fig.chi}
\end{figure}

Seven constraints are imposed on the fitted parameters: four equations 
of the energy-momentum conservation,
two equations for the effective masses 
of the gamma pairs and a required 
intersection of kaon and charged pion trajectories. 
The decay vertex coordinates are not fixed.
The parameters are found by the minimization of the functional with constraints using the method of uncertain Lagrange multipliers and iteration technique. The iterations are stopped when the relative changes of all fitted parameters at the last iteration are less than 10$^{-5}$. For each event all possibilities to associate one of the tracks with charged pion and $\gamma$ pairs with $\pi^{0}$'s are considered. The combination with the least $\chi^{2}$ is used.
Fig.~\ref{fig.chi} shows the $\chi^2$ distributions for the data 
and simulated events. Events with $\chi^2>20$ 
are rejected since in this region the data
exceeds the number of the simulated events due to the high background level. 
Simulation shows that this $\chi^2$ cut reduces statistics of the 
wanted events by 28\% only but decreases background level 
considerably.

\begin{figure}
\begin{center}
\includegraphics*[scale=0.8,bb=0 0 510 510 ]{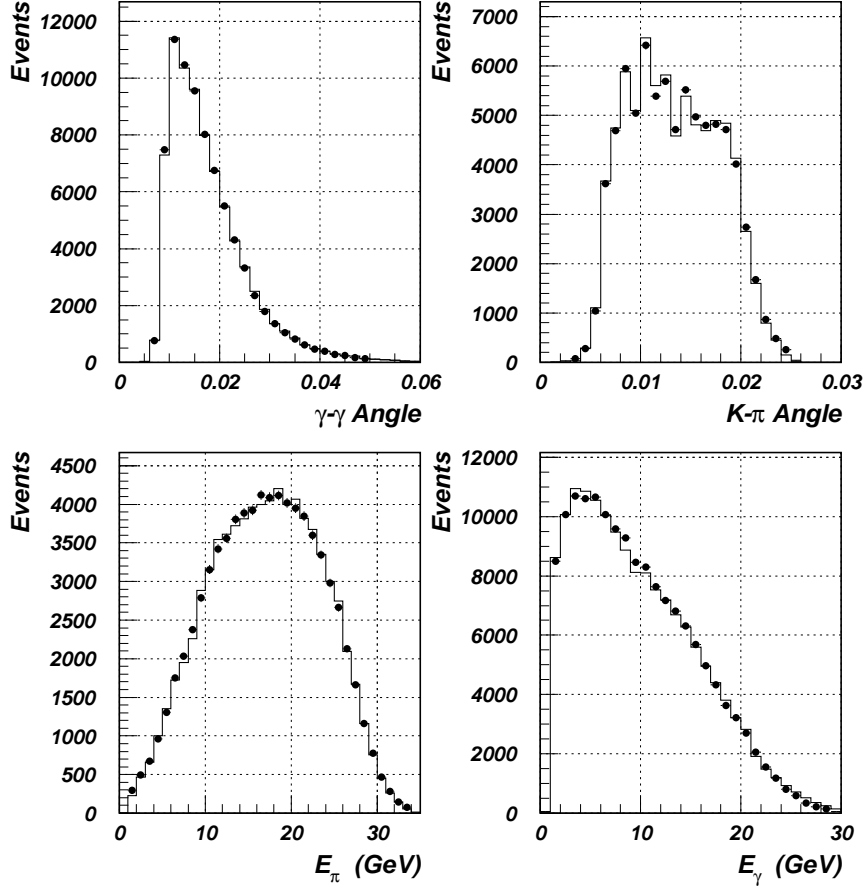}
\end{center}
\caption{
Number of events vs kinematic variables for $K^{\pm}\rightarrow\pi^{\pm}\pi^{0}$
decays (histogram -- simulation, circles -- experiment).
}
\label{fig.cmpmc}
\end{figure}

\begin{figure}
\begin{center}
\includegraphics*[scale=0.8,bb=0 10 510 198]{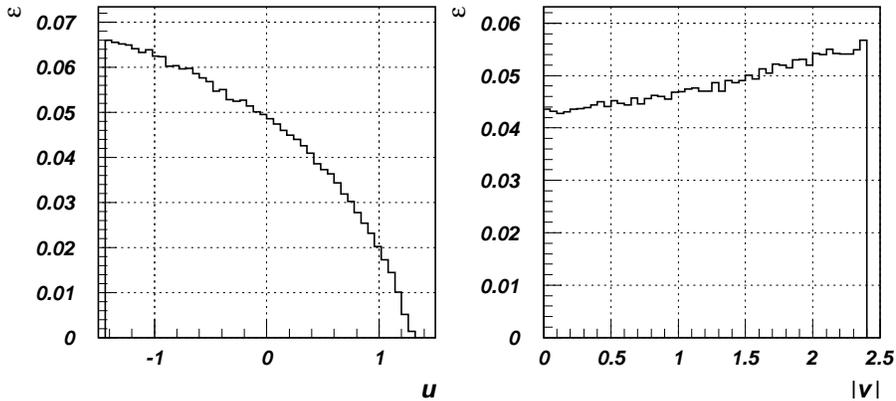}
\end{center}
\caption{
Acceptance vs Dalitz plot variables.
}
\label{fig.accept}
\end{figure}

\begin{figure}
\begin{center}
\includegraphics*[scale=0.8,bb=0 10 510 198]{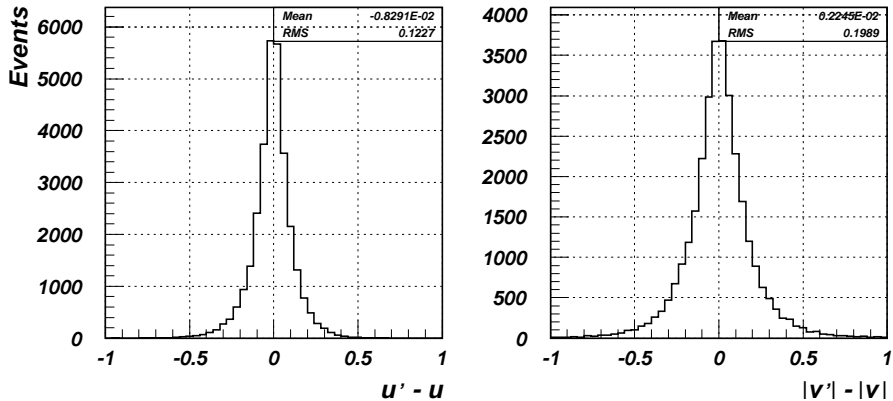}
\end{center}
\caption{
$u,v$ resolutions ($u,v$ -- "true" variables, $u',v'$ -- measured variables).
}
\label{fig.resol}
\end{figure}

The experimental setup operation was simulated using a Monte Carlo (MC) method with the GEANT~3.21 code. 
The setup geometry is described in detail, and the data 
obtained in the experiment were taken into account.
Among these data are the calibration coefficients for each channel of the 
calorimeter, the dependence of the hodoscope efficiency on the particle coordinates and correlations between kaon spatial and angular coordinates 
and its momentum. 
Fig.~\ref{fig.cmpmc} presents different distributions of 
the experimental data and simulated events for 
the $K^\pm \rightarrow \pi^\pm \pi^0$ decays. 
From this figure it follows that there is a good agreement between experimental and simulated distributions. 
Fig.~\ref{fig.accept} shows the acceptance of 
the setup, and 
Fig.~\ref{fig.resol} demonstrates the $u, v$ resolutions averaged over the Dalitz plot. 

\begin{figure}
\begin{center}
\includegraphics*[scale=0.8,bb=0 0 350 250 ]{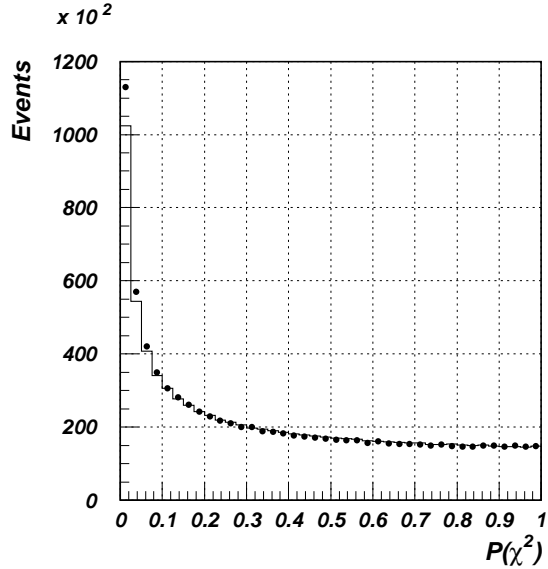}
\end{center}
\caption{
$P(\chi^{2})$ distribution for \kdc\ decays (histogram -- simulation, circles -- experiment).}
\label{fig.pchi}
\end{figure}

The $\chi^2$ probability $P(\chi^2)$ for data and simulation is shown in Fig.~\ref{fig.pchi}. 
Events with $P(\chi^2)>0.1$ are selected for further analysis 
since in this region there is a good agreement 
between the experimental and simulated data.
To check the Level 2 trigger conditions the energies 
corresponding to each of the trigger channels are calculated.
The event is accepted if the number of the channels with energy above
1~GeV is greater than two. This cut rejects only a few \kdc\ events (see the last row of Table 1), but it is important for
$K^{\pm}\rightarrow \pi^{\pm} \pi^{0}$ event selection which is used to 
calibrate the calorimeter, to adjust the simulation code,  and to estimate the 
systematic uncertainties. 

The final data sample is comprised of $N^+ = 278398$ and $N^- = 341015$ events.
Table~\ref{table.select} shows the fraction of events  
rejected by each cut and the cumulative efficiency.

\begin{table}[!h]
\caption{}
\label{table.select}
\bigskip
\begin{tabular}{|c|c|c|}
  \hline
                      & \multicolumn{2}{c|}{Fraction (\%) of events}  
\\ \cline{2-3} 
Selection criteria & rejected by the & passed this and all \\
                   & cut & previous cuts \\
  \hline
$\ge 1$ track  is reconstructed in H1-H3             & 4.4 & 95.6 \\
  \hline
Position of the decay vertex is inside                                                & & \\
the fiducial length of the decay pipe   & 31.2 & 65.7 \\     
  \hline
Number of clusters and tracks  & & \\ 
 corresponds  to the  \kdc\ decay  & 93.2 & 4.48 \\     
  \hline
$\chi^2 < 20$   & 82.2  & 0.80  \\     
  \hline
$P(\chi^2) > 0.1$   & 26.4 & 0.59 \\     
  \hline
  Level 2 trigger is ok & 0.2 & 0.59 \\     
  \hline
\end{tabular}
\end{table}

After all selection cuts for \kdc\ decays, 
an admixture of background remains. 
The background sources are
other modes of kaon decays, 
interactions of beam particles in the material along the beam line 
and overlapping of events due to the finite time resolution of the detectors.  
Simulations of these processes demonstrate that the main contribution to the background comes from the $K^{\pm}\to\pi^{\pm}\pi^0$ (0.21\% ) and $K^{\pm}\to\pi^{\pm}\pi^+\pi^-$ 
(0.03\%) decays. This contribution does not depend on the sign of the 
kaon charge and hence does not cause a false charge asymmetry of the Dalitz plots. The background level from other sources is less than 0.01\%.

The finite energy resolution of the calorimeter results in a noticeable ($\sim$ 10\%) probability of the wrong combinations of 
$\gamma$'s reconstructed into $\pi^0$'s and the hodoscope inefficiency   
can cause a reconstruction of a false track  ($\sim 5\%$). 
Both these effects are taken into account in the event simulation and 
it is found that their influence on the charge asymmetry of the Dalitz plot parameters is negligible.

\section{RESULTS}

\subsection{Difference of the Dalitz plot parameters}

\begin{figure}
\begin{center}
\includegraphics*[scale=0.8,bb=0 10 510 198]{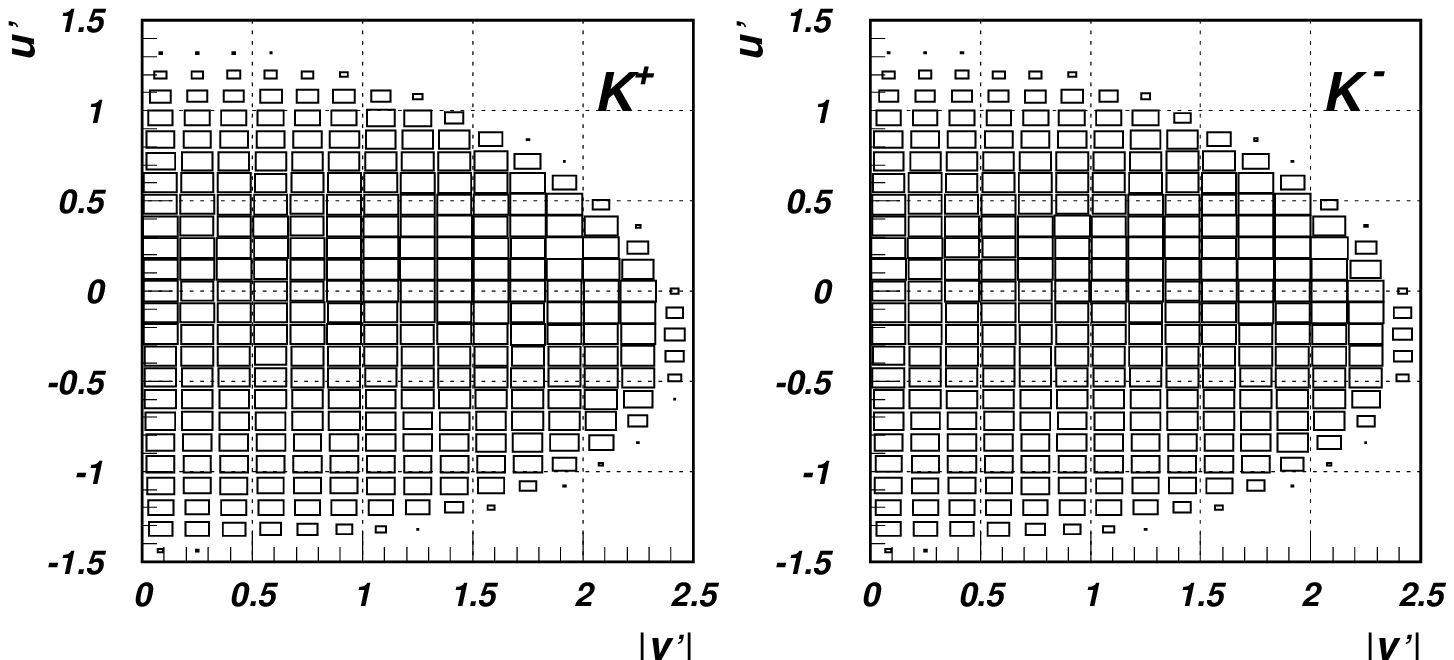}
\end{center}
\caption{
Distribution of events in the Dalitz plots for \kdc\ decays.
}
\label{fig.dalitz}
\end{figure}

\begin{figure}
\begin{center}
\includegraphics*[scale=0.9,bb=50 0 510 198]{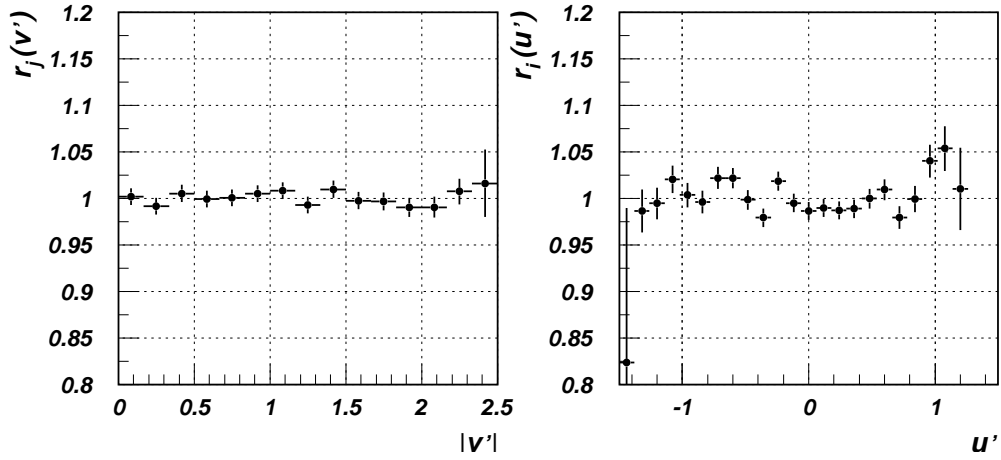} 
\end{center}
\caption{
Ratios of normalized event distributions projected on the $u'$ and $|v'|$ axes.
}
\label{fig.ratiouv}
\end{figure}

\begin{figure}
\begin{center}
\includegraphics*[scale=0.8,bb=0 10 510 510]{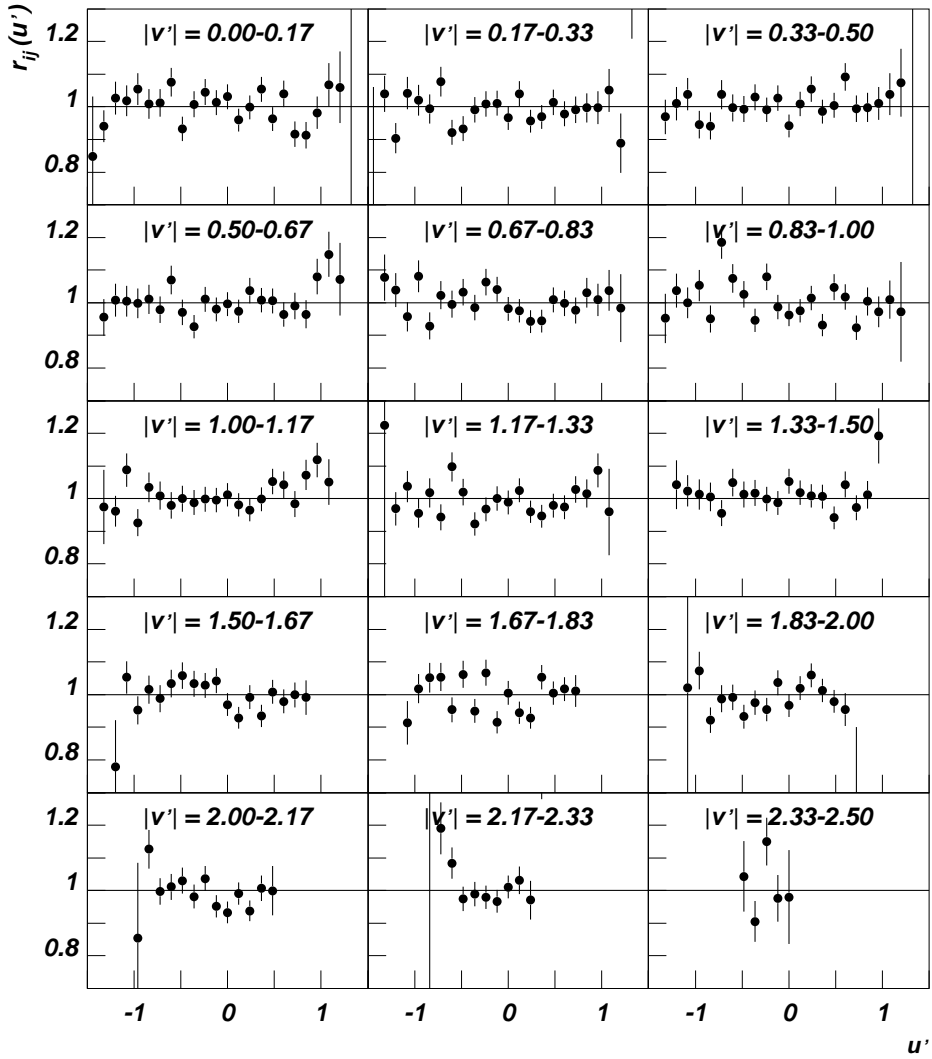}
\end{center}
\caption{
Ratios of normalized event distribution vs $u'$ for the different $|v'|$ intervals.
}
\label{fig.ratio2d}
\end{figure}

The difference of the Dalitz plot parameters
for $K^{\pm}$ decays is estimated by minimizing the following functional form:
\begin{eqnarray}
\chi ^{^2 } (\Delta g,\Delta h,\Delta k) = \sum\limits_{i,j}^{} 
{\frac{{\left( {r_{ij}  - 1 - \alpha_{ij} \Delta g - \beta_{ij} \Delta h - \gamma_{ij} 
\Delta k} \right)^2 }}{{\sigma _{ij}^2 }}}, 
\label{eq.chisq}
\end{eqnarray}

$r_{ij}  = \frac{{{\raise0.5ex\hbox{$\scriptstyle {n_{ij}^ +  }$}
\kern-0.1em/\kern-0.15em
\lower0.25ex\hbox{$\scriptstyle {N^ +  }$}}}}{{{\raise0.5ex\hbox{$\scriptstyle 
{n_{ij}^ -  }$}
\kern-0.1em/\kern-0.15em
\lower0.25ex\hbox{$\scriptstyle {N^ -  }$}}}}$, 
$\sigma _{ij}^2  = r_{ij}^2  \cdot \left( {\frac{1}{{n_{ij}^ +  }} + 
\frac{1}{{n_{ij}^ -  }}} \right)$, where
$n^{\pm}_{ij}$ is the number of events in the $i$-th, $j$-th Dalitz plot bin with 
$u'$, $v'$ measured coordinates (Fig.~\ref{fig.dalitz}), 
and $\alpha_{ij}$, $\beta_{ij}$, and $\gamma_{ij}$ are coefficients defined 
in (A.2) (see Appendix) and calculated by MC.
The values of $\Delta g $, $\Delta h $, and
 $ \Delta k $ as well as the elements of the correlation matrix 
are given in (3):

\begin{equation}
\left\{ \begin{array}{lll}
\Delta g & = & -0.0009 \pm 0.0067,  \\
\Delta h & = & -0.0007 \pm 0.0062, \\
\Delta k & = & -0.0014 \pm 0.0017,
\end{array} \right.
\qquad\qquad\qquad
\left( \begin{array}{ccc}
 1.00 &  0.93 &  0.35 \\
      &  1.00 &  0.32 \\
      &       &  1.00
\end{array} \right)
\end{equation}

The errors shown are statistical only. 
The $\chi^2/{\it ndf}$ is 319/(279-3) = 1.16.

Fig.~\ref{fig.ratiouv} shows the
$r_i(u') =  
\frac{{\sum\limits_j {n_{ij}^ +  } /N^ +  
}}{{\sum\limits_j {n_{ij}^ -  } /N^ -  }}$
and 
$r_j(v') = 
\frac{{\sum\limits_i {n_{ij}^ +  } /N^ +  
}}{{\sum\limits_i {n_{ij}^ -  } /N^ -  }}$
ratios of  the normalized event distributions in Dalitz plots for 
\kdc\ decays projected on the $u'$ and $|v'|$ axes.
The $r_{ij}(u')$  ratio versus $u'$ for different 
$|v'|$ intervals are presented in Fig.~\ref{fig.ratio2d}.

Since some theoretical models 
predict that CP violation in $K^\pm \rightarrow  3\pi$ decays can
be associated with the charge asymmetry
of the slope $g$ only, $\Delta g$ is also estimated assuming 
$ \Delta h  = \Delta k  = 0$ which is in agreement with (3). 
With this assumption we find:
\begin{eqnarray}
\Delta g = 0.0002 \pm 0.0024 \qquad 
\qquad \chi^2/{\it ndf} = 319/(279-1) = 1.15.
\label{eq.resdg}
\end{eqnarray}

The $g$, $h$, and $k$  
parameters appear to be equal for kaons of different signs within statistical uncertainties. But this does not guarantee the 
identity of the event distributions in the corresponding Dalitz plots. 
In order to check the identity of 
$u'$,  $|v'|$, and $(u',|v'|)$ distributions independently of the 
matrix element form (1),  
the Kolmogorov 
nonparametric criterion was used. This analysis provided the following
results: the probabilities that $u'$, $|v'|$ and two-dimensional $(u',|v'|)$  distributions for $K^{+}$ and $K^{-}$ are indistinguishable are equal to
32.4\%, 85.4\% and 55.2\% correspondingly.
The two-dimensional $(u',|v'|)$  distributions were compared using 
the modified Kolmogorov criterion from the 
$HBOOK$  code which works with histograms.

\subsection{Systematic errors}

All measures are taken to assure that $K^{+}$ and $K^{-}$ beams have identical parameters.
Nevertheless the average angles $A_{X}$ and $A_{Y}$ of beam particles with respect to the nominal beam axis and the mean kaon energies in the positive and negative beams could differ by $\Delta A_{X} = 5~\mu rad$, $\Delta A_{Y} = 7~\mu rad$ and
$\Delta E$ = 50~MeV. Simulations show that these uncertainties result in the following systematic errors: 
\[
\delta_A(\Delta g) = 0.0004, \qquad
 \delta_A(\Delta h) = 0.0003, \qquad
 \delta_A(\Delta k) = 0.0001,
\]
\[ 
 \delta_E(\Delta g) = 0.0006, \qquad
 \delta_E(\Delta h) = 0.0004, \qquad
 \delta_E(\Delta k) = 0.0001.
\]

Two methods were used to evaluate systematic uncertainties 
of $\Delta g$, $\Delta h$, and $\Delta k$ connected with $g$, $h$, $k$ errors
(see (2) and (A.1),(A.2) in Appendix): MC and the analytical method assuming
ideal resolutions in $u$ and $v$: $u \equiv u'$, $v \equiv v'$ (see
(A.3),(A.4)). Both methods gave the same results: 

$$
\frac{{\delta (\Delta g)}}{{\Delta g}} \approx
\sqrt {(0.2 \cdot \delta g)^2 + 
(0.6 \cdot \delta h)^2  + (1.6 \cdot \delta k)^2 }, 
$$
$$ 
\frac{{\delta (\Delta h)}}{{\Delta h}} \approx
\sqrt {(0.5 \cdot \delta g)^2 + 
(1.0 \cdot \delta h)^2  + (1.5 \cdot \delta k)^2 }, 
$$
$$
\frac{{\delta (\Delta k)}}{{\Delta k}} \approx
\sqrt {(0.4 \cdot \delta g)^2 + 
(0.5 \cdot \delta h)^2  + (2.9 \cdot \delta k)^2 }. 
$$

Using experimental data \cite{pdg} one can obtain:
\[
\frac{{\delta (\Delta g)}}{{\Delta g}} = 0.014, \qquad 
\frac{{\delta (\Delta h)}}{{\Delta h}} = 0.024, \qquad
\frac{{\delta (\Delta k)}}{{\Delta k}} = 0.019.
\]

Other possible sources of the systematic errors include the time variations of the calorimeter calibration coefficients
and the hodoscope efficiency, the influence of the Earth
magnetic field on the particle beams of different polarity, the difference 
in the $\pi^+$ and $\pi^-$interactions with matter, 
and the difference in composition and intensity of the 
positive and negative beams. The total 
contribution of these factors to the systematic errors does not 
exceed $1\cdot 10^{-4}$. 

We also investigated that varying the minimum energy of $\gamma$'s, the 
minimum and maximum energies of charged pion, the value of the 
$\chi^2$ confidence level 
and the number of reconstructed tracks did not appreciably change the results. The results remain also unaffected
if the bins located at the boundary of the Dalitz plot 
are not used. $\Delta g$, $\Delta h$, and $\Delta k$ can also be found by 
minimizing the functional for the differences of 
the Dalitz plots. 
The obtained results agree with (3) 
and (4).

Thus, the final estimates of the systematic errors are 
\begin{eqnarray}
 \delta(\Delta g) = 7\cdot 10^{-4}, \qquad
 \delta(\Delta h) = 5\cdot 10^{-4}, \qquad
 \delta(\Delta k) = 1.4\cdot 10^{-4}.
\label{eq.syst}
\end{eqnarray}

The systematic errors are approximately an order of magnitude less than 
the statistical errors given in (3).

\section*{CONCLUSIONS}

The differences $\Delta g$, $\Delta h$, $\Delta k$ of  
the Dalitz plot parameters have been measured for the \kdc\ decays 
using the TNF-IHEP facility. The studies were performed in the 35~GeV/c positive and negative hadron beams
at the 70~GeV IHEP accelerator. 
Frequent changes of the beam polarity allow one to minimize the systematic 
uncertainties of the experiment. Our results show that the 
event distributions in the Dalitz plots for $K^{+}$ and $K^{-}$ decays 
are indistinguishable  and that
the $\Delta g$, $\Delta h$, and $\Delta k$ 
are equal to zero within the errors quoted (3),(5).
Assuming $ \Delta h  = \Delta k  = 0$ we find:
\[
\Delta g = 0.0002 \pm 0.0024 (stat.) \pm 0.0007 (syst.).
\]

To find $A_{g} = \Delta g/(g^{+}+g^{-})$ it is assumed $g^+ = g^- = 0.652$ \cite{pdg}: 
\[
A_{g} = 0.0002 \pm 0.0018 (stat.) \pm 0.0005(syst.).
\]

This is the most accurate estimate of the charge asymmetry of the Dalitz plot slope for the \kdc\ decay.

\section* {ACKNOWLEDGMENTS}

We are grateful to A.A.~Logunov, N.E.~Tyurin, 
and A.M.~Zaitzev for their support of 
the experiment; to V.N.~Mikhailin for his assistance in the setup construction and operation; to Yu.V.~Mikhailov, A.N.~Sytin, and V.A.Sen'ko for their help in manufacturing electronics.
We thank the staff of the Accelerator Department and the Beam Division 
who provided high-quality operations of the accelerator complex, 
beam extraction  system, and  the beam channels No.8 and No.23. 
We appreciate the assistance of I.N.~Belyakov, Yu.G.~Nazarov, A.N.~Romadanov, and 
I.V.~Shvabovich in the detector construction. 

This study is supported in part by the Russian Fund for 
Basic Research (grants 02-02-17018, 02-02-17019)
and the President grant 1305.2003.2. 

\newpage
{\large \bf APPENDIX}

\appendix
\section{Calculation of the differences of the Dalitz plot parameters}

In accordance with (1), the probability density function 
can be expressed in the form:  
\[
f(u',v') = \frac{{\int\limits_D {G \cdot (1 + gu + hu^2  + kv^2 )dudv} 
}}{{
\int\limits_D {\int\limits_D {G \cdot (1 + gu + hu^2  + kv^2 )dudvdu'dv'} } 
}},
\]
where $u, v$ and $u', v'$ are  the "true" and measured Dalitz variables,  
and $G \equiv G(u,v,u',v')$ is a function that depends on the efficiencies 
of the detectors and data processing. 
Integration is performed 
over the kinematic boundary of the Dalitz plot. 
This relation can be rewritten as:  
\[
f(u',v') = \frac{{a + gb + hc + kd}}
{{1 + g\bar u + h \overline {u^2 }  + k \overline {v^2 } }},
\]
where 
$ a \equiv a(u',v') = \frac{1}{\varepsilon }\int\limits_D {G\,dudv}$,
$ b \equiv b(u',v') = \frac{1}{\varepsilon }\int\limits_D {u\cdot G\,dudv}$,
$ c \equiv c(u',v') = \frac{1}{\varepsilon }\int\limits_D {u^2\cdot G\,dudv}$,
$ d \equiv d(u',v') = \frac{1}{\varepsilon }\int\limits_D {v^2\cdot G\,dudv}$,
$ \bar u = \frac{1}{\varepsilon }\int\limits_D {\int\limits_D {u \cdot 
G\,dudvdu'dv'} }$,
$ \overline{u^2} = \frac{1}{\varepsilon }
\int\limits_D {\int\limits_D {u^2 \cdot 
G\,dudvdu'dv'} }$, and 
$ \overline{v^2} = \frac{1}{\varepsilon }
\int\limits_D {\int\limits_D {v^2 \cdot 
G\,dudvdu'dv'} }$, and  
$ \varepsilon  = \int\limits_D {\int\limits_D {G\,dudvdu'dv'} } $.
The $ \bar u $, $ \overline{u^2}$, and $ \overline{v^2}$ 
are the mean values of the Dalitz variables and their squares, and
$ \varepsilon $ is the total efficiency of the experiment (including the event reconstruction efficiency) for $|M|^{2}$=1.

Let's introduce the following notations:
\[
g = (g^+ + g^-)/2, \qquad 
h = (h^+ + h^-)/2, \qquad 
k = (k^+ + k^-)/2, 
\] 
\[
 \Delta g = g^+ - g^-, \qquad
 \Delta h = h^+ - h^-, \qquad
 \Delta k = k^+ - k^-.
\] 
Expanding the $\frac{{f^+(u',v')}}{{f^-(u',v')}}$ ratio of 
the normalized Dalitz plots into series in  $\Delta g$, $\Delta h$, 
and  $\Delta k$ and neglecting their quadratic terms,  
we obtain  
\begin{eqnarray}
r(u',v') = \frac{{f^+(u',v')}}{{f^-(u',v')}} \approx 1 + 
{\alpha(u',v')\Delta g + \beta(u',v')\Delta h + \gamma(u',v')\Delta k},
\label{eq.r}
\end{eqnarray}
where 
\begin{eqnarray}
\alpha(u',v') = [b - a\overline u  + h(b\overline {u^2 }  - c\overline u 
) + k(b\overline {v^2 }  - d\overline u )]/D(u',v'), \nonumber \\ 
 \beta(u',v') = [c - a\overline {u^2 } + 
g(c\overline u  - b\overline {u^2 
} )+ k(c\overline {v^2 }  - 
d\overline {u^2 } )]/D(u',v'),  
\\ 
 \gamma(u',v') = [d - a\overline {v^2 } + 
g(d\overline u  - b\overline 
{v^2 } ) + h(d\overline {u^2 }  - c\overline {v^2 } ) ] / 
D(u',v'),\nonumber 
\end{eqnarray}
$$ D(u',v') =
 (1 + g \, \bar u + h\, \overline {u^2 }  + k\, \overline {v^2 } )
  \cdot (a + g \, b + h \, c + k \, d).$$

In the case of $u'\equiv u$ and $v'\equiv v$, ("ideal" resolution) 
formulae (A.1),(A.2) have the form:
\begin{eqnarray}
r_0(u,v) = 1 + 
\frac{{A_0(u,v)\Delta g + B_0(u,v)\Delta h + C_0(u,v)\Delta k}}
{{(1 + g \, \bar u + h\, \overline {u^2 }  + k\, \overline {v^2 } )
 \cdot (1 + g \, u + h \, u^2 + k \, v^2)}},
\label{eq.r0}
\end{eqnarray}
where
\begin{eqnarray}
 A_0(u,v) = u - \overline u  + \,h(u\overline {u^2 }  - u^2\overline u )\,\, 
+ k(u\overline {v^2 }  - v^2\overline u ), \nonumber \\ 
 B_0(u,v) = u^2 - \overline {u^2 }\,\,  + \,g(u^2\overline u  - u\overline 
{u^2 } )\,\, + k(u^2\overline {v^2 }  - 
v^2\overline {u^2 } ),  
\\ 
 C_0(u,v) = v^2 - \overline {v^2 }  + \,
g(v^2\overline u  - u\overline {v^2 } 
)\,\, + h(v^2\overline {u^2 }  - u^2\overline {v^2 } ). \nonumber
\end{eqnarray}

\newpage


\begin{thebibliography}{99}

\bibitem{na48.1} 
A.~J.~Bevan {\it et al.}, Phys. Lett. B465 (1999) 355.
\bibitem{na48.2} 
A.~Lai {\it et al.}, Eur. Phys. J. C22 (2001) 231.
\bibitem{ktev.1} 
A.~Alavi--Harati {\it et al.}, Phys. Rev. Lett. 83 (1999) 22.

\bibitem{pdg} 
K.~Hagiwara et al., Phys. Rev. D66 (2002) 010001.

\bibitem{bel.2} 
A.~A.~Belkov {\it et al.}, Czech. J. Phys. 53 (2003) Suppl. A, \\
hep--ph/0311209.
\bibitem{ambro} 
G.~D'Ambrosio {\it et al.}, Phys. Lett. B273 (1991) 497.
\bibitem{isido} 
G.~Isidori {\it et al.}, Nucl. Phys. B381 (1992) 522.
\bibitem{gamiz} 
E.~Gamiz, hep--ph/0401236 (2004).

\bibitem{istra}
I.~V.~Ajinenko {\it et al.}, Phys. Lett. B567 (2003) 159.  

\bibitem{ford} 
W.~T.~Ford {\it et al.}, Phys. Rev. Lett. 25 (1970) 1370.
\bibitem{smith}
K.~M.~Smith {\it et al.}, Nucl. Phys. B91 (1975) 45.
\bibitem{prelim} 
G.~A.~Akopdzhanov {\it et al.},
in Proceedings of the First International Workshop on Frontier Science ---
 Charm, Beauty, and CP, Frascati, 2002, Ed. by L.~Benussi {\it et al.}
 (LNF, Frascati, 2002), p.~229.

\bibitem{proposal}
V.V. Ammosov {\it et al.}, Preprint IHEP 98-2, Protvino, 1998. 

\bibitem{hodo}
A.V.~Vasiliev {\it et al.}, Instrum.~Exp.~Tech., 1993, vol. 2, p. 50. 

\bibitem{trig} 
Yu.V.~Gilitsky {\it et al.}, Preprint IHEP 93-10, Protvino, 1993. 

\end{thebibliography}
\end{document}